\documentclass[twocolumn,pra,showpacs,preprintnumbers,amsmath,amssymb]{revtex4}

\usepackage{epsfig}

\usepackage{bm}

\begin{document}
\title{On the derivation of amplitude equations for nonlinear oscillators subject to arbitrary forcing}
\author{Catalina Mayol}
\author{Ra\'ul Toral}
\author{Claudio R. Mirasso}
\affiliation{Departament de F\'{\i}sica, Universitat de les Illes
Balears} \affiliation{Instituto Mediterr\'aneo de Estudios
Avanzados (IMEDEA), CSIC-UIB, Ed. Mateu Orfila, Campus UIB, 07122
Palma de Mallorca, Spain} \homepage{http://www.imedea.uib.es}

\date{\today}

\begin{abstract}
By using a generalization of the multiple scales technique we develop a method to derive amplitude equations for zero--dimensional forced
systems. The method
allows to consider either additive or multiplicative forcing terms and can be
straightforwardly applied to the case that the forcing is white noise. We give
examples of the use of this method to the case of the van der Pol--Duffing
oscillator. The writing of the amplitude equations in terms of a Lyapunov
potential allow us to obtain an analytical expression for the probability
distribution function which reproduces reasonably well the numerical simulation
results.
\end{abstract}

\pacs{
02.30.Mv, 
05.10.-a, 
05.10.Cg, 
05.40.Ca. 
}
\keywords{Amplitude equations. Multiple scale method. Additive noise. Multiplicative noise. Van der
Pol--Duffing oscillator.}

\maketitle

\section{Introduction}

It is known that the general trends of the behavior of a dynamical system can be
captured, in some cases of interest, by the so--called amplitude equations
\cite{nayfeh,graham96}, describing the slow dynamics of the envelope of the
trajectories. One of the most interesting features of amplitude equations is their
universality: many systems can share the same amplitude equation depending only on
general symmetry considerations.  Another advantage of using the amplitude equations
is that they allow numerical integration with a bigger integration step. A very
successful technique, amongst others, for the derivation of amplitude equations is
that of the  multiple scales method. This technique has been extensively applied, for
instance, to both unforced and periodically forced nonlinear oscillators
\cite{nayfeh,nayfeh2,nayfeh3}. However, for more general, e.g. non periodic, forcing
terms there is no systematic derivation of the amplitude equation. In this paper we
develop a possible extension of the multiple scales method to  obtain amplitude
equations for dynamical systems forced with general functions. As an application, we
consider oscillators which are randomly forced  either by additive noise
\cite{proctor,daff,scal,devries,falk} or by multiplicative noise
\cite{daff,geno,arno1,arno2}.

The derivation of amplitude equations for randomly forced dynamical systems has
been previously considered in the literature. In some works, dynamical systems
are treated in a probabilistic way and noise effects on bifurcations are
studied \cite{holmes,kushe2,kushe,stone}. A different approach has been
used in references \cite{drolet2001,blom} where the authors derive a stochastic
Landau form as the amplitude equation for the stochastic Swift--Hohenberg model
and use it to describe the dynamics of the bifurcating solutions. A simplifying
feature of this case is that the Swift--Hohenberg equation has only first order
derivatives in time. Second order equations have been studied in \cite{drolet}
at the level of probability distribution functions. Our method, being an
straightforward extension of the multiple scales method, can be easily applied
to dynamical systems including second order time derivatives.

This paper is structured as follows: In section \ref{sec1}, the general theory
is presented using the van der Pol--Duffing oscillator with an arbitrary
forcing term as an example. In section \ref{sec2}, the same  oscillator is
considered under the influence of an additive noise forcing term. In section
\ref{sec3}, we present the results for the multiplicative noise case. Finally,
the main conclusions of this paper are reviewed in section \ref{sec4}.

\section{General theory}
\label{sec1}

Although we believe our method to be quite general, for the sake of
concreteness, in this section we study an oscillatory system, namely a van der
Pol--Duffing oscillator, with a general additive forcing term, as defined by
the following equation for the variable $x(t)$:
\begin{equation}
\ddot{x}+x=\epsilon ~[k_1 (1- x^2)~\dot{x} - k_2 x^3] ~+ \epsilon
~f(t), \label{eq1}
\end{equation}
being $f(t)$ any time dependent function and $\epsilon$ is considered as a
small parameter. The specific cases $k_1=1, \,k_2=0$ (van der Pol oscillator)
and  $k_1=0, \, k_2=1$ (Duffing oscillator) are contained in this general
equation. For $f(t)=0$, the van der Pol--Duffing oscillator has an unstable
fixed point and a stable limit cycle in the  phase space $(x, \dot{x})$.
Equation parameters have been rescaled out such that the frequency of the
linear oscillator, i.e. $\epsilon=0$, is $\omega_0=1$. For $\epsilon > 0$, the
evolution can be written as $x(t)=A(t) e^{i\omega_0 t}+\bar A(t)e^{-i\omega_0t}$, being $A(t)$
the slowly varying complex  amplitude ($\bar{A}$ denotes the complex conjugate of $A$).

We extend the method of multiple scales in order to be able to
consider equation (\ref{eq1}) for a general function $f(t)$. Let
us briefly review the method of multiple scales\cite{nayfeh}. In
this method, one looks for a series expansion of the time
dependent variable, $x(t)$, of the form:
\begin{equation}
x(t)=x_0(t)+\epsilon x_1(t)+\epsilon^2 x_2(t)+\dots. \label{eq2}
\end{equation}
The main point is to consider different time scales: $T_0$, $T_1$, $T_2$,...,
with $T_m=\epsilon^m t$ as independent variables and hence any function of time becomes a function of the $T_m$'s:
\begin{eqnarray}
x_0(t) & = & x_0(T_0,T_1,\dots)\\ x_1(t) & = &
x_1(T_0,T_1,\dots)\\ & \vdots & \\ f(t) & = & f(T_0,T_1,\dots).
\end{eqnarray}
The time derivative is transformed according to
\begin{equation}
\frac{d}{dt}= \frac{\partial}{\partial T_0} + \epsilon \frac{\partial}{\partial T_1} +
 \epsilon^2 \frac{\partial}{\partial T_2} + \dots =
  D_0 +\epsilon D_1 + \epsilon^2 D_2 + \dots,
\end{equation}
where $D_i \equiv \frac{\partial}{\partial T_i}$. By substituting
this expansion into (\ref{eq1}) and equating coefficients up to
order $\epsilon^1$, one obtains:
\begin{eqnarray}
 (D_0^2+1) x_0 &  =&  0, \label{eq3}\\
(D_0^2+1) x_1& =& k_1 (1-x_0^2) D_0 x_0 - k_2 x_0^3 - 2 D_0 D_1
x_0 + \nonumber\\& &  f(T_0, T_1,\dots).\label{eq4}
\end{eqnarray}

The solution of (\ref{eq3}) is
\begin{equation}\label{eq5}
x_0 (T_0,T_1,\dots)= A(T_1) e^{i T_0}+ \bar{A}(T_1)
e^{-i T_0}
\end{equation}
where it has
been assumed that the amplitude $A(T_1)$ of the sinusoidal
solution depends only on the slow time variable $T_1$ instead of
all the sequence $T_1,T_2,\dots$. Accordingly, $f(t)=f(T_0,T_1)$.
Replacing this solution in (\ref{eq4}), we get:
\begin{eqnarray}
(D_0^2+1) x_1& =& [i~k_1 A (1 -  |A|^2) - 2 i A'- 3 k_2 A |A|^2 ]
e^{i T_0} \label{eq6}\nonumber\\\nonumber
 & &+ [-i k_1 \bar{A}( 1 - |A|^2) + 2 i \bar{A}'-3 k_2  A |A|^2] e^{-i T_0}
 \\\nonumber
 & &  - i A^3 e^{3 i T_0} + i \bar{A}^3 e^{-3 i T_0}+f(T_0, T_1)\\
& \equiv&
g(T_0, T_1),
\end{eqnarray}
where $A'$ denotes the derivative of  $A$ with respect to $T_1$, and we define
 $g(T_0, T_1)$ as
the r.h.s. of the equation.
The amplitude equation is obtained by avoiding resonant terms with the frequency of
the l.h.s of the equation,
\begin{equation}
\langle e^{iT_0}|g(T_0,T_1) \rangle=0,
 \label{eq7}
\end{equation}
with the scalar product for the $T_0$ variable defined as
\begin{equation}
\langle v(T_0)|w(T_0) \rangle= \int_{-\infty}^{\infty}
\bar{v}(T_0) w(T_0) dT_0,
\end{equation}
and the result $\langle e^{inT_0}|e^{imT_0}\rangle=\delta_{n,m}$.
By using  (\ref{eq6}) and (\ref{eq7})  one obtains
\begin{equation}
i [ k_1 A (1-|A|^2)-2 A'] - 3 k_2 A |A|^2
 + \langle e^{i T_0}|f(T_0,T_1) \rangle =0.
\label{eq8}
\end {equation}
The problem has been reduced to extract from $f(t)=f(T_0,T_1)$
the resonant terms, i.e. those with a component of $e^{iT_0}$. These terms are those giving a contribution
different from zero in the previous equation.

The particular case of a periodic forcing term, $f(t)=\cos(\lambda ~t)$, with
a forcing frequency $\lambda= \omega_0+\sigma \epsilon=1+\sigma
\epsilon$, and $\sigma = {\cal O}(1)$ (soft resonant excitation)
has been extensively studied before \cite{nayfeh}. In this case,
the standard approach considers the decomposition:
\begin{eqnarray}
\cos [\left(1 + \sigma \epsilon)t \right)] & = & \cos(t + \sigma \epsilon
t)= \cos(T_0+\sigma T_1) = \nonumber \\ & & \frac{1}{2}\left( e^{iT_0} e^{i\sigma
T_1}+e^{-iT_0} e^{-i\sigma T_1}\right).
\end{eqnarray}
When substituting this expression in Eq. (\ref{eq8}) for the $A$ variable, the
term multiplying $e^{iT_0}$ is the only one giving a non-zero contribution. In other words, the contribution to the
equation for $A$ is the spectral component of the function $f(t)$
at the frequency $\omega_0=1$. The resulting {\sl amplitude equation} is:
\begin{equation}
\frac{d A}{d T_1}= k_1
\frac{1}{2}A (1-|A|^2) + i k_2 \frac{3}{2}A|A|^2 - \frac{i}{4}
e^{i \sigma T_1}. \label{eq9}
\end{equation}
It is clear that the proper time scale for the variation of the amplitude $A$
is given by $T_1$.

For any other function $f(t)$ the splitting of the time
$t$ in terms of slow, $T_0$, and fast, $T_1$, variables might
not be so straightforward. In general we will have to compute expressions of the form:
\begin{equation}\label{eq10}
\langle e^{i T_0}| e^{i m T_0} f(T_0,T_1) \rangle.
\end{equation}
Our proposal is to make the following decomposition:
\begin{equation}
f(T_0,T_1)=e^{-(m-1)iT_0} e^{i(m-1) T_1/\epsilon} f(T_1/\epsilon),
\label{eq11}
\end{equation}
which has the property that the contribution
to the scalar product (\ref{eq10}) is different from zero:
\begin{equation}\label{eq12}
\langle e^{i T_0}| e^{i m T_0} f(T_0,T_1) \rangle=
 e^{i(m-1) T_1/\epsilon} f(T_1/\epsilon).
\end{equation}
This simple substitution rule has to be applied with the necessary values of $m$ as demanded in each case. Notice that this decomposition gives always a non vanishing contribution to the scalar product. This is particularly interesting in the case of nearly constant functions $f(t)$ for which the spectrum has a peak at
$\omega=0$, far from the main peaks at $\pm\omega_0$. In the white noise case, our proposal is able to extract a suitable contribution from the flat spectrum.

\begin{figure}
\begin{center}
\epsfig{file=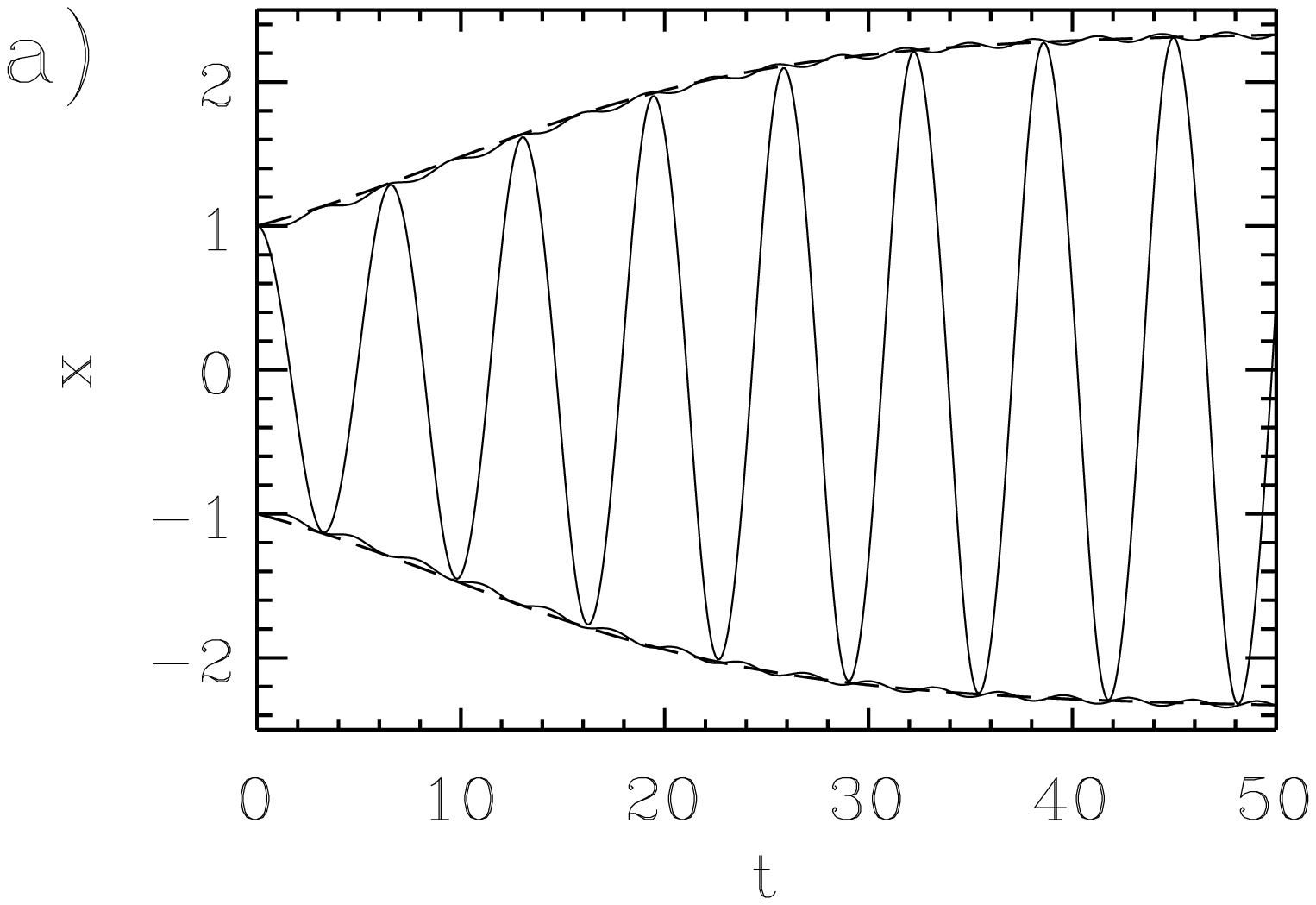,width=6cm}
\epsfig{file=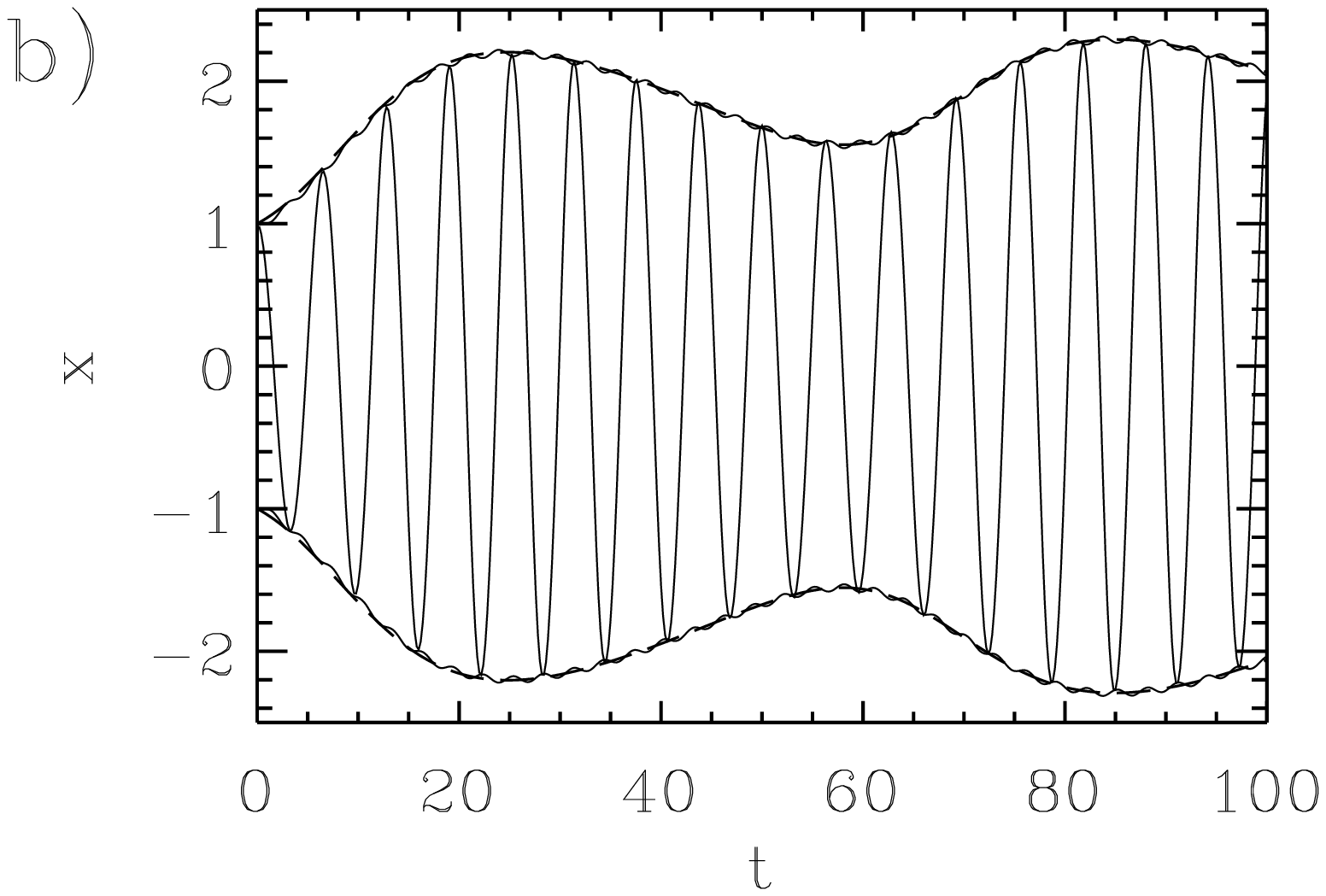,width=6cm}
\epsfig{file=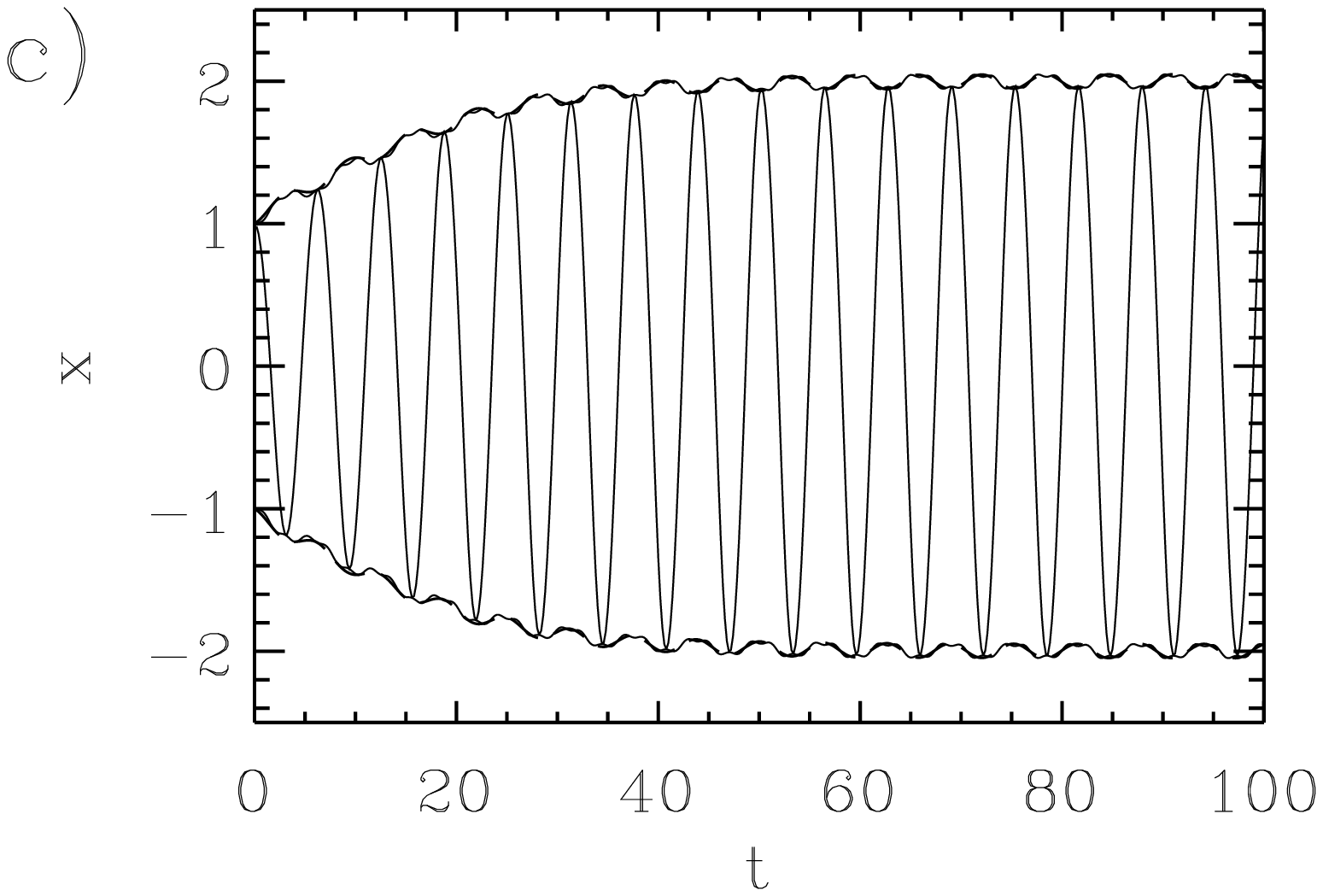,width=6cm}
\epsfig{file=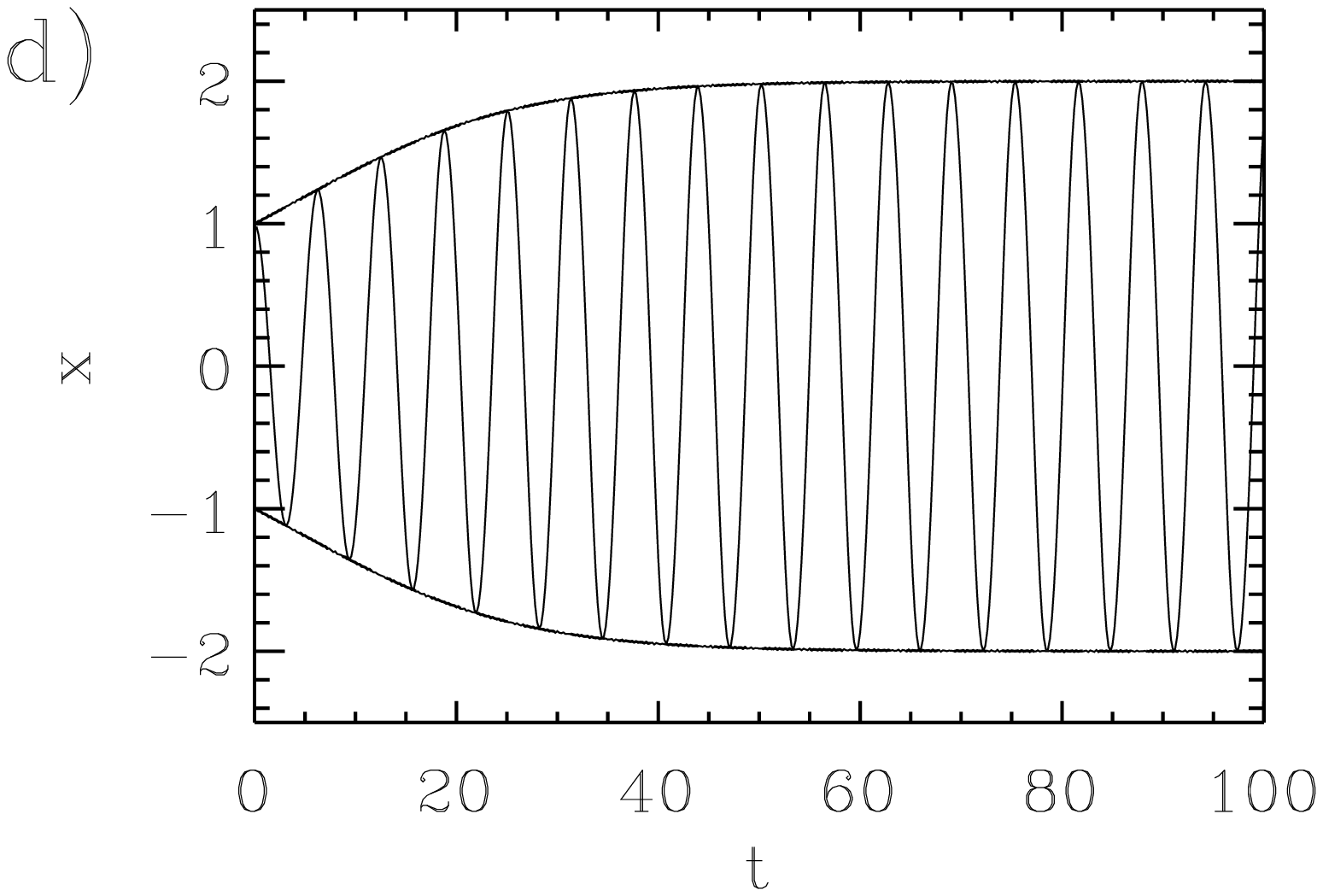,width=6cm}
\caption{\label{lfig1} Time evolution of $x(t)$ in a van der Pol oscillator  (\ref{eq1}) (oscillating thin solid line) forced with $f(t)=\cos(1+\sigma t)$. 
In order to extract the envelope from the complex amplitude $A$ we
also plot $\pm 2 R$ where $A= R e^{i\phi}$ in two cases: the solid line is the result of method (\ref{eq13a}), the dashed line comes from the standard method (\ref{eq9}). Notice that both methods give almost indistinguishable results. The values of the parameters are $\epsilon=0.1$,  $k_1=1$, $k_2=0$. The initial conditions: $x(0)=1$, $\dot{x}(0)=0$ are used henceforth in all the examples given. The different plots are: a) $\sigma=0$; b) $\sigma=1$; c) $\sigma=10$; d) $\sigma=500$.}
\end{center}
\end{figure}

Coming back to our example, Eq. (\ref{eq8}) contains a term of the form (\ref{eq10}) with $m=0$. Substitution of the ansatz Eq.(\ref{eq12}) yields:
\begin{equation}
\frac{d A}{d T_1} =k_1\frac{A}{2}
(1 - |A|^2) +i k_2 \frac{3}{2}A|A|^2- \frac{i} {2} e^{-i T_1/\epsilon} f(T_1/\epsilon).
\label{eq13}
\end{equation}
This is our result for the amplitude equation of the van der Pol-Duffing
oscillator under general forcing $f(t)$. Concretely, the choice $f(t)= \cos[(1+\sigma\epsilon) t]$ gives:

\begin{equation}
\frac{d A}{d T_1} =k_1\frac{A}{2}
(1 - |A|^2) +i k_2 \frac{3}{2}A|A|^2- \frac{i} {4} e^{i \sigma T_1} - \frac{i} {4} e^{-i \sigma T_1}e^{-i 2T_1/\epsilon}.
\label{eq13a}
\end{equation}
As compared to the standard result of Eq. (\ref{eq9}), this equation includes  an extra term that gives oscillations in the amplitude at the scale $T_1/\epsilon$. It is true that in this case these oscillations belong to the fast time scale $T_0$ and could, in principle, be eliminated thus leading to the standard result, Eq. (\ref{eq9}). However, as shown in figure \ref{lfig1}, the effect of this extra term is very small and decreases with increasing $\sigma$.

\begin{figure}
\begin{center}
\epsfig{file=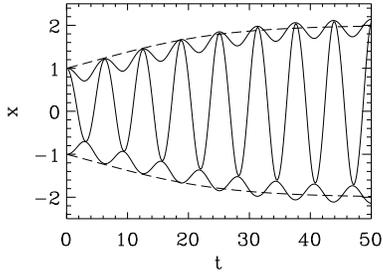,width=6cm} \caption{\label{lfig2} Time evolution of $x(t)$ in a van der Pol oscillator (\ref{eq1}) (oscillating thin solid line) forced with $f(t)=2.0$. We also plot the envelope $\pm 2  R(t)$ as coming from Eq. (\ref{eqb13}) (solid line). The parameters are $\epsilon=0.1$, $k_1=1$, $k_2=0$. Notice that the fit of the envelope to the trajectory of $x(t)$ worsens if we neglect the last term in Eq. (\ref{eqb13}), as shown by the dashed line, specially in the negative values of $x(t)$.}
\end{center}
\end{figure}

We now consider the forcing term $f(t)=f_0$, constant, for which the standard method does not obtain any contribution. Our method immediately yields:
\begin{equation}
\frac{d A}{d T_1} =k_1\frac{A}{2}
(1 - |A|^2) +i k_2 \frac{3}{2}A|A|^2- \frac{i} {2} e^{-i T_1/\epsilon} f_0.
\label{eqb13}
\end{equation}
One could again decide to eliminate the oscillations in the fast scale $T_1/\epsilon$. However, as shown in figure \ref{lfig2}, the influence of the extra term can actually improve upon the predictions of the amplitude equation. In particular, it can describe the lack of symmetry around the zero value observed in the evolution of $x(t)$.

The above scheme can be applied to the case of multiplicative forcing term.
Again, we consider a specific example: the van der Pol--Duffing equation with a linear multiplicative term of the form
\begin{equation}
\ddot{x}+x=\epsilon ~[ k_1(1- x^2)~\dot{x} - k_2 x^3] ~+ \epsilon
x~f(t). \label{eq14}
\end{equation}
The multiple scales ansatz (\ref{eq2}) leads in the first order approximation to the form (\ref{eq5}). The next order yields
\begin{eqnarray}
& & D_0^2 x_1+x_1= \nonumber\\
& & k_1 (1-x_0^2) D_0 x_0 - k_2 x_0^3 -2 D_0 D_1 x_0
+x_0 f(T_0,T_1) \nonumber\\ \nonumber
    &=&e^{i T_0}[-2 i \frac{\partial A}{\partial T_1} +i k_1 A - i k_1 A |A|^2
     - 3 k_2 |A|^2 A+ \\\nonumber & &f(T_0,T_1) A]\\\nonumber
    &+& e^{-i T_0} [2 i \frac{\partial \bar{A}}{\partial T_1}
    -i  k_1 \bar{A}  +i k_1 \bar{A} |A|^2  -3 k_2 |A|^2
     \bar{A} +  \\\nonumber & & f(T_0,T_1) \bar{A}]  \\\nonumber
    &-& e^{3 i T_0} A^3 (i k_1+ k_2) + e^{-3 i T_0} \bar{A}^3 (i k_1-k_2)
    \\\label{eq14b}
    &\equiv &g_1(T_0,T_1).
\end{eqnarray}
The nullity of the scalar product $\langle e^{i T_0} | g_1(T_0,T_1) \rangle=0$
involves two contributions of the form (\ref{eq10}): one with $m=1$ and another with $m=-1$. These contributions are computed according to the general rule
(\ref{eq12}). The amplitude equation obtained is
\begin{eqnarray}
\frac{d A}{d T_1}&  = &  k_1 \frac{1}{2} A (1-|A|^2) +i k_2 \frac{3}{2}
A |A|^2 - \frac{i}{2} A f(T_1/\epsilon) \nonumber \\ & &- \frac{i}{2} \bar{A}
e^{-2 i T_1 /\epsilon} f(T_1 /\epsilon). \label{eq15}
\end{eqnarray}

\begin{figure}
\begin{center}
\epsfig{file=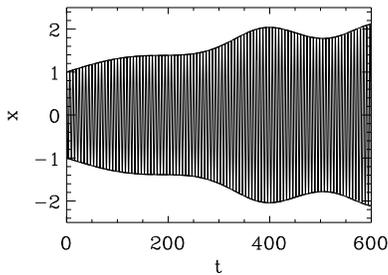,width=6cm} \caption{\label{lfig3} Time evolution of $x(t)$ in a van der Pol--Duffing oscillator (\ref{eq14}) (oscillating thin solid
line) with multiplicative forcing $x f(t)=x \cos(2 t)$, together with the envelope $\pm 2 R(t)$ coming both from (\ref{eq15a}) and (\ref{eq15b})
(they are indistinguishable at the scale of the figure). The values of the parameters are $k_1=1$, $k_2=1$, $\epsilon=0.01$.}
\end{center}
\end{figure}

Let us be more specific and consider $f(t)=\cos(2t)$. The amplitude equation is:
\begin{eqnarray}
\frac{d A}{d T_1} & =&  k_1 \frac{1}{2} A (1-|A|^2) +i k_2 \frac{3}{2}
A |A|^2 \nonumber \\ & & -\frac{i}{4} \bar A - \frac{1}{2} \bar{A}\cos{\left(\frac{2T_1}{\epsilon}\right)} -\frac{i}{4}\bar A e^{-4 i T_1 /\epsilon}. \label{eq15a}
\end{eqnarray}
Again the last two terms, belonging to the temporal scale $T_0$, could be
discarded reobtaining the amplitude equation derived after replacing directly
 $\cos(2t)=\cos(2T_0)$ in Eq. (\ref{eq14b}):
\begin{equation}
\frac{d A}{d T_1} = k_1 \frac{1}{2} A (1-|A|^2) +i k_2 \frac{3}{2}
A |A|^2 - \frac{i}{4} \bar A . \label{eq15b}
\end{equation}
In figure \ref{lfig3} we compare the results of Eqs. (\ref{eq14}), (\ref{eq15a}) and (\ref{eq15b}). Observe that both amplitude equations faithfully follow the envelope of the $x(t)$ variable.

\section{Additive noise}
\label{sec2} We apply the above developed technique to the case of an additive noise forcing term, while in a following
section we will consider multiplicative noise. We take equation
(\ref{eq1}) with a real noise term $f(t)=\xi(t)$ with zero mean
value and temporal correlations
 $\langle \xi(t) \xi(t') \rangle = \delta(t-t')$.
The general amplitude equation (\ref{eq13}) adopts now the form:
\begin{equation}
\frac{d A}{d T_1} =k_1\frac{A}{2}
(1 - |A|^2) +i k_2 \frac{3}{2}A|A|^2- \frac{i} {2} e^{-i T_1/\epsilon} \sqrt{\epsilon}\xi(T_1).
\label{eq13n}
\end{equation}
with $\langle \xi (T_1) \xi (T_1') \rangle = \delta(T_1-T_1')$,
and it has been used that $\xi(T_1/\epsilon)=\sqrt{\epsilon} \,
\xi(T_1)$.
The statistical properties that follow from this equation are contained in the corresponding Fokker--Planck
equation for the time evolution of the probability density
function $P(A,t)$ \cite{risken}. It is possible to obtain an approximate analytical expression in the steady state $P_{st}(A)$. We first
write down the real and imaginary parts of the amplitude equation
\begin{eqnarray}
\frac{d A_r} {d T_1} & = &\frac{k_1}{2} A_r [ 1- (A_r^2 +A_i^2)] -k_2
\frac{3}{2} A_i (A_r^2+A_i^2)\nonumber \\ & & - \frac{1}{2} \sin(T_1/\epsilon)
\sqrt{\epsilon}~ \xi(T_1),\\ 
\frac{d A_i} {d T_1} & = & \frac{k_1}{2}
A_i[ 1- (A_r^2 +A_i^2)]+ k_2 \frac{3}{2} A_r (A_r^2+A_i^2) \nonumber \\ & & -
\frac{1}{2} \cos(T_1/\epsilon) \sqrt{\epsilon}~ \xi(T_1),
\label{eq16}
\end{eqnarray}
with $A=A_r+i A_i$. It can be shown that the deterministic terms are that of a non-relaxational potential flow with the Lyapunov function\cite{maxi,lyapunov}:
\begin{equation}
V(A)=k_1[-\frac{|A|^2}{4}+\frac{|A|^4}{8}]
\end{equation}
Using this fact, one can obtain an approximate analytical expression for the stationary probability distribution $P_{st}(A)$. To this end, we simplify the stochastic set of equations (\ref{eq16}) by substituting the cosine and sine functions
by its root mean square value $\langle \sin^2(t)\rangle^{1/2}=\langle \cos^2(t)
\rangle^{1/2}=1/\sqrt{2}$ and considering  that $\langle \sin(t)
\cos(t)\rangle=0$, effectively introducing different noise terms for  the real and imaginary parts. This yields:
\begin{eqnarray}
\frac{d A} {d T_1} = \frac{k_1}{2} A [ 1- |A|^2] + i  k_2
\frac{3}{2}A |A|^2 -\frac{1}{2} \sqrt{\frac{\epsilon}{2}}~
\eta(T_1), \label{eq17}
\end{eqnarray}
where $\eta$ is a complex noise term  with  zero mean value and
correlations $\langle \eta(T_1)~ \bar{\eta}(T_1') \rangle =
\delta(T_1-T_1')$.
It turns out that this new noise terms satisfy the fluctuation-dissipation relation\cite{maxi}, and the stationary probability distribution is given in terms of the Lyapunov potential as
\begin{equation}
P_{st}(A) \propto \exp(-V(A)/\tilde{\epsilon}),
\label{eq18}
\end{equation}
where $\tilde{\epsilon}=\epsilon/16$.

\begin{figure}
\begin{center}
\epsfig{file=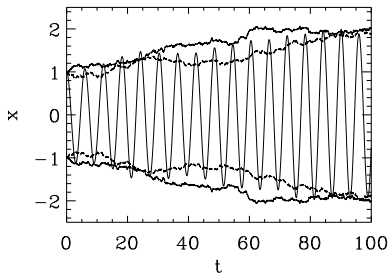,width=6cm} \caption{\label{lfig4} Time evolution of $x(t)$ in a van der Pol--Duffing oscillator (\ref{eq1}) (oscillating thin solid
line) forced with additive white noise $f(t)=\xi(t)$, and the envelope $\pm 2 R(t)$ coming from (\ref{eq16}) (solid line) and from the 
approximated expression (\ref{eq17}) (dashed line). The parameters are $k_1=1$, $k_2=1$. $\epsilon=0.1$. Notice that the individual trajectories are not exactly approximated by the amplitude equations, while the mean values would be (see next figure).}
\end{center}
\end{figure}

\begin{figure}
\begin{center}
\epsfig{file=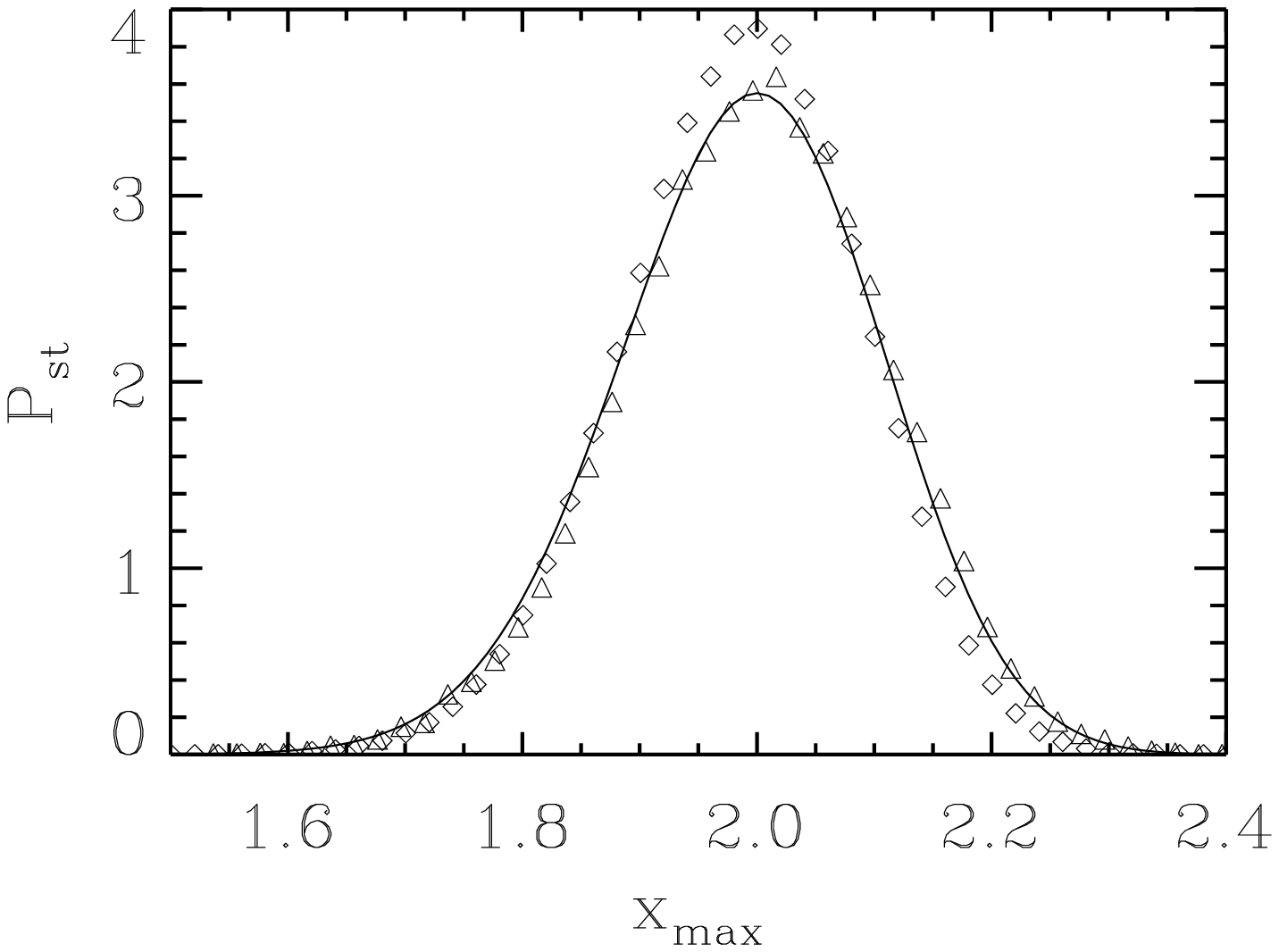,width=6cm} \caption{\label{lfig5} The diamonds correspond to the histogram for the maxima of $x$ for a van der
Pol--Duffing oscillator in the same case of additive white noise forcing and parameters than in the previous figure \ref{lfig4}. The triangles are the histogram of the envelope $2  R(t)$ as obtained from the amplitude equation (\ref{eq16}).  The solid line is the probability distribution function as given by the approximate analytical result of Eq. (\ref{eq18}). }
\end{center}
\end{figure}

We have simulated the dynamics of Eq. (\ref{eq1}) with
$f(t)=\xi(t)$ following standard stohcastic integration methods as given in \cite{maxi}.
The results  are compared with the corresponding amplitude
equation, i.e. (\ref{eq16}), in Fig. \ref{lfig4}. We observe that
the amplitude equation does not fit the actual maxima $x_{max}$ of $x(t)$ for a given
realization. However, by taking  many realizations, the averaged amplitude
does fit the maxima of the average values. As an evidence, we show in Fig.
\ref{lfig5}, the histogram for the $x_{max}$ is compared to the
histogram for the amplitude. In solid line, it appears the
approximate theoretical expression (\ref{eq18}). The concordance
between  solid line and  triangles reveals the validity of  the approximation
which allowed us to derive (\ref{eq17}) from (\ref{eq16}).

\section{Multiplicative noise}
\label{sec3}

\begin{figure}
\begin{center}
\epsfig{file=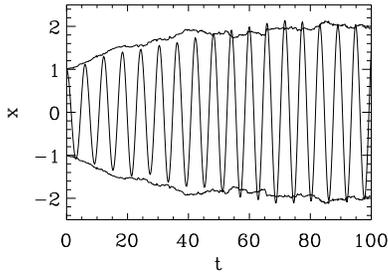,width=6cm} \caption{\label{lfig6} Time evolution $x(t)$ for a van der Pol--Duffing oscillator (thin solid oscillating line) with a multiplicative noise term $x f(t)=x \xi(t)$,  and the envelope $\pm 2 R(t)$ coming from the amplitude equation (\ref{eq15mn}) (solid
line). The parameters are $k_1=1$, $k_2=1$,
$\epsilon=0.1$.}
\end{center}
\end{figure}

In this section we consider the van der Pol oscillator with a
multiplicative noise term of the form $f(t)=x\xi(t)$. The corresponding amplitude equation (\ref{eq15}) becomes:
\begin{eqnarray}
\frac{d A}{d T_1} & =&  k_1 \frac{1}{2} A (1-|A|^2) +i k_2 \frac{3}{2}
A |A|^2 \nonumber \\ & & - \frac{i}{2} A \epsilon^{1/2}\xi(T_1) - \frac{i}{2} \bar{A}
e^{-2 i T_1 /\epsilon} \sqrt{\epsilon}\xi(T_1). \label{eq15mn}
\end{eqnarray}
In this  case of multiplicative noise terms,  the time traces are
similar to the ones observed for additive noise terms, see Fig.
\ref{lfig6}.

\begin{figure}
\begin{center}
\epsfig{file=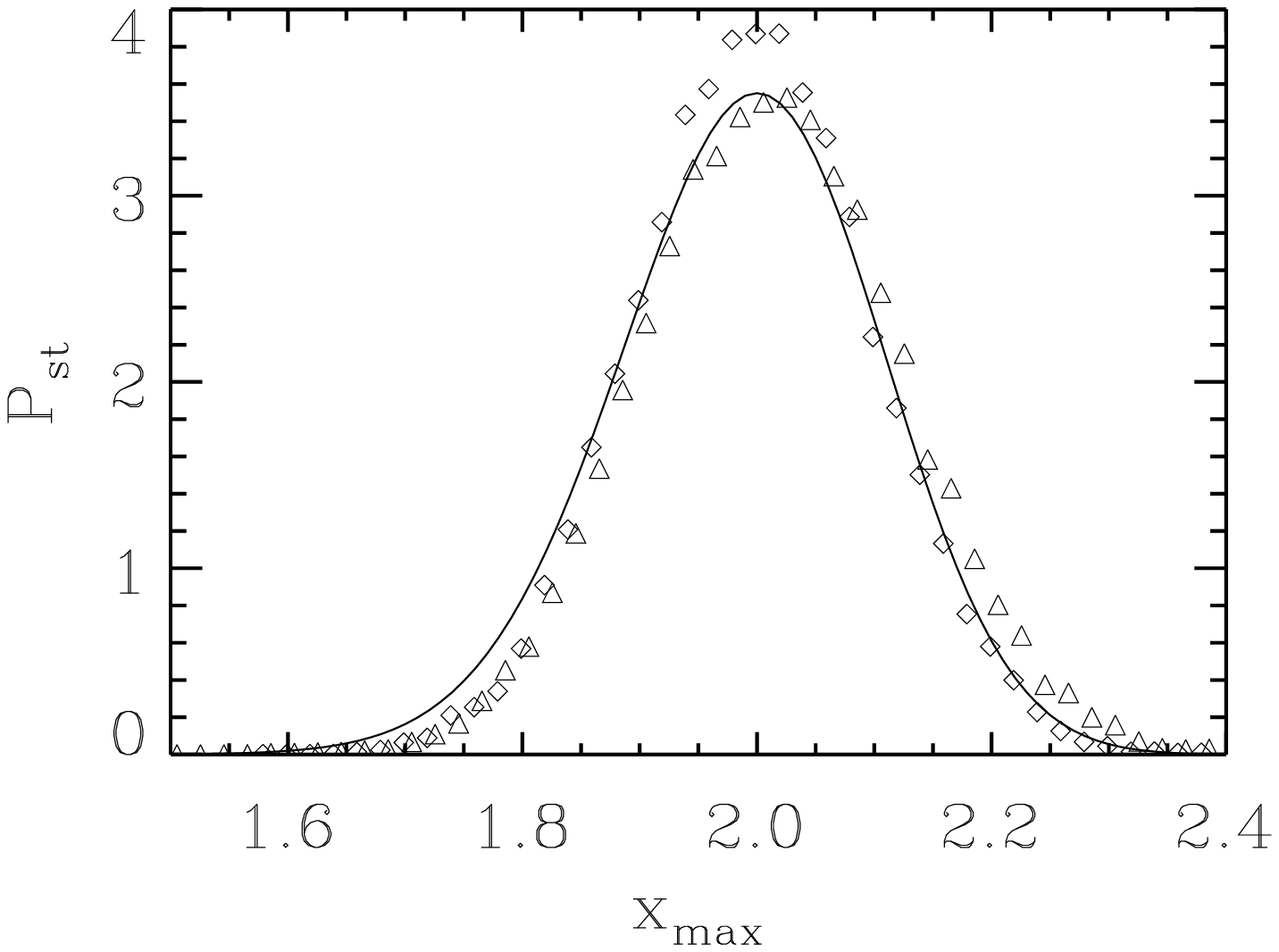,width=6cm} \caption{\label{lfig7} The diamonds correspond to the histogram for the maxima of $x$ for a van der
Pol--Duffing oscillator in the same case of multiplicative white noise forcing and parameters than in the previous figure \ref{lfig5}. The triangles are the histogram of the envelope $2  R(t)$ as obtained with the amplitude equation (\ref{eq15mn}).  The solid line is the probability distribution function as given by the approximate analytical result of Eq. (\ref{lasteq}). }
\end{center}
\end{figure}

In order to be able to compute the stationary probability distribution, we now introduce modulus and phase variables: $A=R \,e^{i \phi}$, the resulting equations are
\begin{eqnarray}\nonumber
\frac{d R}{d T_1} & = & k_1 \frac{1}{2} R (1 -R^2) \\ & & -\frac{1}{2}
 R \, \xi(T_1/\epsilon) \sin(2[T_1 /\epsilon+\phi]),\\
\frac{d\phi} {d T_1} & = & k_2 \frac{3}{2} R^2 -\frac{1}{2}
 \xi(T_1 /\epsilon) \\ & &-\frac{1}{2} \, \xi(T_1/\epsilon)
 \cos(2[T_1 / \epsilon+\phi]).
\end{eqnarray}
Again, we approximate the sine and cosine terms by its root mean square
value. In this way, the equation for the modulus does not contain the phase variable:
\begin{eqnarray}
\frac{d R}{d T_1} & = & k_1 \frac{1}{2} R (1- R^2)-\frac{1}{2} R
\sqrt{\frac{\epsilon}{2}} \xi(T_1). \label{eq19}
\end{eqnarray}
This equation can be analyzed using the general methods \cite{maxi, graham1, graham2} and it can be shown that, despite the fact that the Lyapunov function does not satisfy the fluctuation dissipation relation, Eq.(\ref{eq18}) still holds asymptotically in the limit $\epsilon \rightarrow 0$ giving:
\begin{equation}
\label{lasteq}
P_{st}(R)\propto  \exp\left(-\frac{4k_1}{\epsilon}[-R^2+\frac{R^4}{2}]\right).
\end{equation}
This result is compared in Fig. \ref{lfig7} with the numerical simulations.

We would like to end this section by commenting that equation
(\ref{eq19}) has the same structure that Eq. (35) of reference
\cite{drolet}, in the sense that the noise term  $\xi(T_1)$
appears multiplying the variable $R$. In that reference, the
result is obtained by working at the level of the probability
distribution function. Our method, which we believe to be simpler
and more straightforward, works directly at the level of the
(stochastic) dynamical equation.

\section{Conclusions}
\label{sec4}

In this work, we develop a simple, general method to obtain amplitude equations for a
large variety of dynamical systems forced by an arbitrary time dependent function,
$f(t)$, including noises, both additive and multiplicative. Our method is based on
the multiple scales analysis combined with a recipe to extract the resonant terms for
arbitrary forcing functions. As a representative example, we analyze in detail the
van der Pol--Duffing oscillator.

In the deterministic case, our method is able to reproduce standard results. It only
differs from them in the presence of some extra terms in the amplitude equations. We
show that those terms are either very small or improve upon the predictions of the
equation. For example, it can incorporate the asymmetries observed in the
simulations, as in Fig. 2.

In the stochastic additive case, $f(t)=\xi(t)$, our method is able to satisfactorily capture its contribution to the amplitude equation. A simple set of approximations allow us to obtain analytically the stationary probability distribution $P_{st}(A)$ in terms of a Lyapunov potential function.

Similar conclusions can be drawn for the multiplicative noise, $f(t)=x \xi(t)$. Again, our method allows us to obtain easily the stochastic amplitude equation. An approximation, valid now in the limit of small noise, yields the stationary distribution.

\section{Acknowledgments}

Fruitful discussions with Jorge Vi\~{n}als are gratefully acknowledged. We wish
to thank Thomas Erneux for useful comments. We acknowledge financial support
from MCyT (Spain) and FEDER under projects BFM2001-0341-C02-01 and BMF2000-1108.

\newpage

\end{document}